\documentstyle[12pt]{article}

\begin{document}
\hfill{LA-UR-92-3478}

\vspace{7pt}
\begin{center}
{\large\sc{\bf New Look at QED$_4$: the Photon as a
Goldstone Boson and }}

{\large\sc{\bf the Topological Interpretation of Electric
Charge.}}
\baselineskip=12pt
\vspace{20pt}

A. Kovner$^ *$\\
\vspace{10pt}
Theory Division, T-8,
Los Alamos National Laboratory,
MS B-285\\
Los Alamos, NM 87545\\
\vspace{5pt}
 and \\
\vspace{5pt}
B. Rosenstein$^{**}$  \\
\vspace{10pt}
Institute of Physics,
Academia Sinica\\
Taipei 11529.,
Taiwan, R.O.C.\\
\vspace{20pt}
\end{center}
\begin{abstract}
We develope the dual picture for Quantum Electrodynamics in
3+1 dimensions.  It is shown that the photon is massless in
the Coulomb phase due to spontaneous breaking of the
magnetic symmetry group. The generators of this group are
the magnetic fluxes through any infinite surface $\Phi_S$.
The order parameter for this symmetry breaking is the
operator $V(C)$ which creates an infinitely long magnetic
vortex. We show that although the order parameter is a
stringlike rather than a local operator, the Goldstone
theorem is applicable if $<V(C)>\ne 0$. If the system is
properly regularized in the infrared, we find $<V(C)>\ne 0$
in the Coulomb phase and $V(C)=0$ in the Higgs phase. The
Higgs - Coulomb phase transition is therefore understood as
condensation of magnetic vortices. The electric charge in
terms of $V(C)$ is topological and is equal to the winding
number of the mapping from a circle at spatial infinity into
the manifold of possible vacuum expectation values of a
magnetic vortex in a given direction.  Since the vortex
operator takes values in $S^1$ and $\Pi_1(S^1)={\cal Z}$,
the electric charge is quantized topologically.
\end{abstract}

\vspace{10pt}
\
\newline
*KOVNER@PION.LANL.GOV
\newline
**BARUCH@PHYS.SINICA.EDU.TW
\vfill
\pagebreak

\section{Introduction}
Gauge theories play a dominant role in modern elementary
particle physics. It is clear beyond reasonable doubt that
all the interactions of elementary particles known to date
are described by a gauge theory.  As a consequence in some
physicists minds gauge symmetry attained a status of a
philosophical principle.  It must be noted however, that the
reason for this is purely empirical.  The "gauge principle"
does not have the same deep philosophical roots as, say the
equivalence principle of general relativity or the
uncertainty principle of quantum mechanics. Mainly this is
because it pertains to the form of description, the
"language" in which one describes a physical law rather than
to the physical law itself. This language proved to be
indispensible when formulating renormalizable theories of
interacting vector particles. In many instances it is also
very convenient for actual calculations since the degrees of
freedom used in this description are almost free and the
perturbation theory can be easily applied. Such is the case
in $QED$ and the ultraviolet region of QCD.

 In some cases however, although a neat mathematical
construction, the gauge symmetry in fact obscures rather
than highlights the underlying nonperturbative physics.  The
main conceptual difficulty with the gauge description is
that it makes use of redundant nonphysical quantities which
often makes the interpretation of a calculation almost as
difficult as performing the calculation itself.

 One example of such a situation is the understanding of the
(constituent) quark degrees of freedom in QCD. On the large
nonphysical Hilbert space those appear as multiplets of the
``global color'' $SU(3)$ group. However this group acts
trivially on the physical gauge invariant states of the
theory.  Hence the problem of understanding in physical
terms what is precisely meant by the color and its
confinement.

 It is of course possible in principle to fix the gauge
completely.  However in a completely gauge fixed formulation
the fields that appear in the Lagrangian are as a rule
nonlocal. For example in QED, fixing axial gauge turns the
matter fields into variables localized on a line rather than
at a point \begin{equation}
\phi_{axial}(x)=\phi(x)\exp\{ie\int_{x}^{\infty}
dy^3A^3(y)\}
\end{equation}
where the initial fields $\phi(x)$ and $A_\mu(x)$ are
``local'' but on the nonphysical Hilbert space. In the
Coulomb gauge the matter field \begin{equation}
\phi_{coulomb}(x)=\phi(x)\exp\{ie\int d^3y
A^i(y)\frac{x^i-y^i}{|x-y|^3}\} \end{equation} creates the
electric field of a point charge and has therefore a
nonvanishing support everywhere in space. When written in
terms of these fields, the Lagrangian is nonlocal and the
theory looks very different from a local field theory which
would usually please one's eye.

Because of this unfortunate feature there are several
interesting physical questions already in the simplest,
abelian gauge theories which either do not arise naturally
or tend to be ignored in the framework of the standard gauge
description.  Here are several of these, which motivated us
in the present research.

1. Exact masslessness of a photon. In the standard
formulation the masslessness of a photon is almost a
consequence of kinematics.  However, we know that the photon
is not always massless in the Higgs model, and that this is
in fact a profound dynamical effect. When one discovers a
massless particle the natural question is: what keeps it
massless. The simplest possible explanation is that it is
the Goldstone theorem. So there is a question whether the
photon in QED a Goldstone boson and if yes of what symmetry.

2. The nature of the Higgs - Coulomb phase transition. The
Higgs - Coulomb phase transition is usually described as due
to spontaneous breaking of the electric charge in the Higgs
phase. This description is however not without a flaw.
Electric charge, being equal to a surface integral of the
electric field at spatial infinity does not have a local
order parameter. (This is the reason why the Goldstone
theorem is not applicable in the Higgs phase.) Consequently
the Coulomb and the Higgs phase differ only in expectation
values of {\it nonlocal} fields. In this situation however,
there is no physical argument that tells us that the two
phases must be separated by a phase transition.  In fact in
the similar situation in $Z_N$ gauge theories it is known
that the phases are analytically connected
\cite{szlachanyi}.  In QED, however the two phases are
separated by a genuine phase transition which is second or
first order depending on the values of parameters. The
question is whether one can give a different, complementary
characterization of the Higgs and Coulomb phases in QED
which will make clear that those are really distinct phases.
Usually such an explanation involves spontaneous breaking of
some {\it global} symmetry.

3. Topological nature of the electric charge. The electric
current in QED is trivially conserved: $\partial_ \mu
J^\mu=\partial_\mu (\partial_\nu F^{\mu\nu})=0$.  In the
quantum theory the charge is also quantized. Both these
features would automatically follow if the electric charge
could be represented as a topological charge associated with
nontrivial homotopy of a vacuum manifold. The possibility
that the electric charge could be topological is not so
unnatural. One can measure the charge by making local
measurements of electric field far from it making use of
Gauss law. For a nontopological charge this would be
impossible. Maybe it is possible to find in QED a set of
variables in terms of which the degeneracy of the vacuum and
the topological nature of the electric charge are explicit.

It would be very interesting to find an alternative
formulation of a gauge theory, or at least (since the exact
reformulation turns out to be very difficult) an alternative
basis in which these questions become natural and the
answers to them are relatively straightforward.

In fact in 2+1 dimensions there exists a ``dual''
representation that allows to answer all of these questions
in the affirmative.  There one is able to define such a
variable: the complex vortex operator $V(x)$ \cite{marino},
\cite{npb}. Although it is defined in terms of an
exponential of a line integral of the electric field, it can
be shown that it is a local scalar field
\cite{local}. It is an eigenoperator of the conserved charge - the
magnetic flux through the plane \cite{ezawa}.  In the
Coulomb phase the field $V(x)$ has a nonvanishing
expectation value and the flux symmetry is spontaneously
broken \cite{npb},\cite{preskill}. This results in the
appearance in the spectrum of a massless Goldstone boson -
the photon.  The electric charge when expressed in terms of
the vortex field has the form of a topological charge
associated with the homotopy group $\Pi_1(S^1)$:
$Q\propto\int d^2x
\epsilon_{ij}\partial_i(iV^*\partial_j V +c.c.)$
\cite{prl}.
 In the Higgs phase $<V>=0$. Consequently the charges are
completely screened and there is no massless particle in the
spectrum.

 In this picture it is clear that the Higgs and the Coulomb
phases must be separated by a genuine phase transition line.
The relevant symmetry, the magnetic flux symmetry, is
restored in the Higgs phase. The Nielsen-Olesen (NO)
vortices exist in this phase as particles that carry the
corresponding charge: .  The Coulomb - Higgs phase
transition can be thought of as condensation of the NO
vortices in the Coulomb phase.  Moreover on the basis of
universality one concludes, that whenever it is second
order, the Higgs - Coulomb phase transition must be in the
universality class of the XY model.

In 1+1 dimensional QED the dual representation also exists.
Since in 1+1 dimensions there is no massless photon and no
Coulomb - Higgs phase transition, only the third question is
can be asked. In this case the electric charge can be
represented as topological in terms of the dual field
$\sigma$: $Q=\int dx
\partial\sigma=\sigma(+\infty)-\sigma(-\infty)$. For the massless and
massive Schwinger model the standard bosonization procedure
leads to exact reformulation of the theory in terms of the
field $\sigma$ only\cite{coleman1} and thereby to the {\it
exact} dual Lagrangian. In the scalar Higgs model, the dual
transformation can be only performed approximately
\cite{rabinovich}, but the topological interpretation of the
electric charge is nevertheless achieved.

 The aim of this paper is to develop a similar picture for
the 3+1 dimensional Higgs model.  In Section 2 we discuss
the analog of the flux symmetry in 3+1 dimensions.  The
conserved currents of this magnetic symmetry are the
components of the dual field strength tensor $\tilde
F_{\mu\nu}$.  Because of the constraint $\partial_iB_i=0$ no
local order parameter can be found. The 3+1 dimensional
analog of $V(x)$ are stringlike operators $V(C)$ which
create infinitely long magnetic vortex lines along a curve
$C$.

In section 3 we show that although these operators are not
local they are still good order parameters, in the sence
that the Goldstone theorem is applicable in the phase where
they have a nonzero expectation value. We also show that in
this phase the electric charge is topological in terms of
$V(C)$ and is thereby quantized.  An electrically charged
state carries a unit wind of the phase of $V(C_{\vec u})$
where $V(C_{\vec u})$ is the set of all magnetic vortices
associated with straight lines in the direction $\vec u$.
In the classical approximation, indeed $V(C)\ne 0$ in the
Coulomb phase.

Due to the infinite length of the vortex line $V(C)$ can not
have a finite expectation value beyond classical
approximation. Infrared divergences due to phase
fluctuations of $V(C)$ lead to vanishing of the VEV even in
the Coulomb phase, not unlike vanishing of an order
parameter in 1+1 dimensional theories with a ``classically
broken'' continuous symmetry.  In the present case however
one can define a regularized model in which one of the
spatial dimensions is compact. Vortex lines parallel to this
direction will then have finite expectation value in the
Coulomb phase and the Goldstone theorem and the topological
interpretation of the electric charge will be retained. This
is discussed in Section 4.

Section 5 is devite to a brief discussion of the dual
picture and possible extension of this approach to
nonabelian theories.

\section{The magnetic symmetry, the Coulomb - Higgs phase
transition and the vortex operator in 3+1 dimensions.}

Approaches to all the three questions mentioned in the
introduction which we are addressing in this paper have one
common element.  They all require a construction of a
sufficiently local (gauge invariant) order parameter.  Let
us briefly recall how this was constructed in 2+1
dimensions.

The symmetry which is broken in the Coulomb phase of the 2+1
dimensional Higgs model is the magnetic flux symmetry
generated by
\begin{equation}
\Phi=\int d^2x B(x)
\end{equation}
with the conserved current $\tilde F_\mu$.  The defining
relation for the vortex operator $V(x)$ therefore was the
commutation relation with the magnetic field
\begin{equation}
[B(x),V(y)]=-g\delta^2(x-y)V(y)
\end{equation}
One also insisted on the locality of $V(x)$: it had to
commute with all gauge invariant local fields at space -
like separations.
\begin{equation}
[V(x),O(y)]=0,\ \ x\ne y
\end{equation}
These conditions determined $V(x)$ up to a multiplicative
local gauge invariant factor as
\begin{equation}
V(x) = C \exp\{\frac{i}{e}\int
d^2y[\epsilon_{ij}{(x-y)_j\over
(x-y)^2}E_i(y)+\Theta(x-y)J_0(y)]\}
\label{oprmdef}
\end{equation}

where $\Theta (x)$ is an angle between the vector $x_i$ and
the $x_1$ axis, $ 0 < \Theta < 2 \pi$.  The requirement of
locality lead in particular to a quantization condition on
possible eigenvalues of the magnetic flux $g$: $g=\frac
{2\pi n}{e}$

The operator $V(x)$ has a simple physical meaning: it
creates a pointlike magnetic vortex of the strength $g$. In
the Higgs phase the magnetic flux symmetry is not broken and
the NO vortices behave like particles. In the Coulomb phase
they condence. This breaks the flux symmetry spontaneously
and leads to the appearance of the massless photon.

\subsection{The vortex operator.}
Let us now try to implement the same program in the 3+1
dimensional Higgs model.

The analog of the conserved flux current in 3+1 dimensions
is the dual magnetic field strength tensor $\tilde
F_{\mu\nu}$.  Its conservation equation is again fust the
homogeneous Maxwell equation of electrodynamics. It was
shown in \cite{ijmp} that the matrix element of $\tilde
F_{\mu\nu}$ between the vacuum and the one photon state in
the Coulomb phase has the characteristic form of a matrix
element of a spontaneously broken current between the vacuum
and a state with one Goldstone boson. In the circular
polarization basis
\begin{equation}
<0|\tilde F_{0i}(0)|e^\lambda_{\pm}, {\vec p}>=
\mp {i\over (2\pi)^{3/2}} \sqrt{p_0\over 2} e^{\pm}_i
{\rm lim}_{p^2\rightarrow 0}{1\over 1-\Pi(p^2)}
\end{equation}
\begin{equation}
<0|\tilde F_{ij}(0)|e^\lambda_{\pm}, {\vec p}>=
-{i\over (2\pi)^{3/2}} \sqrt{p_0\over 2}\epsilon_{ijk}
e_{\pm}^k{\rm lim}_{p^2\rightarrow 0}{1\over 1-\Pi(p^2)}
\end{equation}
where $\Pi(p^2)$ is the vacum polarization.

One can define many conserved charges associated with the
currents $\tilde F_{\mu\nu}$. In particular the magnetic
flux through any infinite surface S
\begin{equation}
\Phi_S=\int_S dS^i B_i
\label{phis}
\end{equation}
is time independent. Since the magnetic flux through any
closed surface vanishes, the set of independent $\Phi_S$ is
given by the set of boundaries (at spatial infinity) rather
than the set of surfaces themselves.  It will suffice for
our purposes however to consider only the planes
perpendicular to the three coordinate axes. We define the
magnetic charge $\Phi_i$ as the average magnetic flux
through a plane perpendicular to the i-th axis

\begin{equation}
\Phi_i\equiv \lim_{L\rightarrow\infty}\frac{1}{2L}\int_{-L}^L d x^i
\int_{S(x^i)} d\vec S\vec B(x^i)
\label{phii}
\end{equation}
where $S(x^i)$ is the plane perpendicular to the i-th axis
with the i-th coordinate $x^i$.

We now construct an order parameter for $\tilde F_{0i}=B_i$.
This is an operator which creates magnetic vortices. Clearly
in 3+1 dimensions no local operator of this kind can be
constructed. Since the magnetic field is divergenceless, the
magnetic flux must either form closed loops or infinitely
long lines.  Closed loops however do not carry any global
charge.  The best one can do therefore is to construct an
operator creating an infinite vortex line. This operator
should be "line - local".

Any gauge invariant local field should commute with $V(C)$
at all points but on the line $C$.  In particular
\begin{equation}
[B_i(x),V(C)]=gl_i(x,C)V(C), \ \ l_i(x,C)=\int d\tau
\delta(x-\bar x(\tau))\frac{d\bar x_i(\tau)}{d\tau}
\end{equation}
The commutator of $V(C)$ and $J_i(x)$ should also vanish for
$x\not\in C$.  Analogously to 2+1 dimensions, these two
conditions determine $V(C)$ as
\cite{polchinski}
\begin{equation}
V_n(C)=\exp\left\{i\frac{n}{e}\int
d^3y[a_i(y-x)E_i(y)+b(x-y)J_0(y)]\right\}
\end{equation}
where $a_i(x)$ is a vector potential of an infinitesimally
thin magnetic vortex along $C$ : $\epsilon_{ijk}\partial_j
a_k(x)=l_i(x,C)$ and the function $b(x)$ satisfies
$\partial_i[b(x)]_{mod 2\pi}= a_i(x)$. Since $a_i(x)$ has a
nonvanishing curl, the function $b(x)$ must have a surface
of singularities ending at the curve $C$. For example, for a
straight line $C$ running along the $x_3$ axis one has
$a_i(x)=\epsilon_{ij}\frac{x_i}{x_1^2+x_2^2},\ \ i=1,2;\ \
a_3(x)=0$ and $b(x)=\Theta(x)$ with $\Theta$ the polar angle
in the $x_1-x_2$ plane (Fig.1.).
\begin{figure}
\caption{ The function $\Theta(x)$.}
\label{f1}
\end{figure}
As in 2+1 dimensions, the operator $V(C)$ is an operator of
a singular gauge transformation with the gauge function
$nb(x)$. This ensures the commutativity of $V(C)$ with any
local gauge invariant operator outside the line of
singularities $C$. The single valuedness of the gauge
transformation imposes the quantization condition on
possible fluxes in a vortex $g=\frac{2\pi n}{e}$. This of
course corresponds to the well known fact that the Abrikosov
vortices in the Higgs phase carry quantized flux. From now
on we will concentrate on the elementary vortex operator
$n=1$.

Choosing $C_i$ as a straight line parallel to the i-th axis
we have
\begin{equation}
[V(C_i), \Phi_j]=\delta_{ij}\frac{2\pi}{e}V(C_i)
\label{vi}
\end{equation}
Using the Gauss' law and integrating by parts one can recast
the vortex operator into the following form
\begin{equation}
V(C)=\exp\{i \frac{2\pi}{e} \int_{S: \partial S=C} dS^iE^i\}
\label{vor}
\end{equation}
where the integration is over the half plane bounded by $C$
(Fig.2).
\begin{figure}
\caption{ The vortex operator
 $V(C_3)$ which creates the magnetic flux tube parallel to
the third axis.}
\label{f2}
\end{figure}

\subsection{The vortex operator and the dual vector potential.}

Let us now digress a little bit and show how the vortex
operator can be represented in terms of the dual vector
potential. This is particularly simple in the case of a free
photon. The Gauss's law in this case reduces to
$\partial_iE_i=0$ and can be solved by introducing the dual
vector potential $\chi_i$ via
\begin{equation}
E_i=\epsilon_{ijk}\partial_j\chi_k
\label{chi}
\end{equation}
To reproduce the correct equal time commutation relations we
must also have
\begin{equation}
B_i(x)=\pi_i(x)
\end{equation}
where $\pi_i$ is canonically conjugate to $\chi_i$.  Of
course the field $\chi_i$ is determined by eq.(\ref{chi})
only up to a gradient of a scalar function.  The
transformation
\begin{equation}
\chi_i\rightarrow\chi_i+\partial_i\lambda
\label{dgt}
\end{equation}
is generated by $\partial_iB_i$ and is in fact a magnetic
gauge symmetry due to the constraint
\begin{equation}
\partial_iB_i=0
\label{dgl}
\end{equation}
As in the case of the electric gauge symmetry, one should
however be careful. The transformations eq.(\ref{dgt}) with
gauge functions $\lambda$ that satisfy
$\lim_{x\rightarrow\infty}\lambda(x)\rightarrow 0$ are
indeed generated by $\exp \{i\int
\lambda(x)\partial_iB_i(x)\}$ and are therefore gauge
symmetries. However if the function $\lambda$ does not
vanish somewhere at the spatial infinity, the transformation
eq.(\ref{dgt}) is generated by $\exp \{i\int
\partial_i(\lambda B_i)\}$. The operator of the
transformation is not a unit operator on the constraint
eq.(\ref{dgl}) and the transformation therefore is a true
physical symmetry.  So, for example if
$\lambda(x)=a\delta^2(x-x(V_S))$ where $x(V_S)$ are points
inside a half space bounded by the surface $S$, one has a
global transformation generated by $\Phi_S$ of
eq.(\ref{phis}).  The magnetically gauge invariant operators
are the t'Hooft's loops \cite{thooft1}(the dual analogs of
Wilson loops) or infinite t'Hooft's lines
\begin{equation}
V(C)=\exp\{ig\int_Cdl_i\chi_i\}
\label{vchi}
\end{equation}
In a theory of a free photon the constant $g$ is not
quantized.

In the interacting theory the Gauss's law is
\begin{equation}
\partial_iE_i-J_0=0
\end{equation}

We can now define the dual potential in the following way.
Since the electric current is conserved it can be
potentiated
\begin{equation}
J_\mu=\epsilon_{\mu\nu\lambda\rho}\partial_\nu K_{\lambda\rho}
\end{equation}
The tensor potential $K_{\mu\nu}$ is defined up to a Kalb - Ramond

gauge transformation
\begin{equation}
K_{\mu\nu}\rightarrow K_{\mu\nu}+\partial_{[\mu}M_{\nu]}
\label{kalb}
\end{equation}
Let us fix this Kalb - Ramond gauge symmetry by requiring
that for any surface $S$
\begin{equation}
\int_S dS^i\epsilon_{ijk}K_{jk}=en(S)
\end{equation}
where $e$ is the electric coupling constant and $n(S)$ is an
integer which depends on the surface. In QED this is an
admissible gauge fixing, since the divergenceless part of
$\epsilon_{ijk}K_{ij}$ can be changed arbitrarily by a
choice of $M_i$ and the charge inside any closed surface is
an integer of $e$.  The dual vector potential is then
defined by \begin{equation}
E_i-\epsilon_{ijk}K_{jk}=\epsilon_{ijk}\partial_j\chi_k
\end{equation}
With this definition one has
\begin{equation}
\exp\{i\frac{2\pi n}{e}\int_Cdl_i\chi_i\}=\exp\{i\frac{2\pi
n}{e}\int_{S:\partial S=C}dS^iE^i\}
\end{equation}
Comparing this with eq.(\ref{vor}) we find
\begin{equation}
V(C)=\exp\{i\frac{2\pi}{e}\int_Cdl_i\chi_i\}
\end{equation}

\section{The Goldstone
theorem and the topological interpretation of the electric
charge. The classical approximation.}
\subsection{The Goldstone theorem.}
We have now constructed eigenoperators of the magnetic
symmetry in 3+1 dimensions.  The first natural question to
ask is whether their vacuum expectation value vanish. First
let us consider the classical approximation.  We will
calculate the quantum corrections in the next section.
Although in the infinite system they change the results in a
very important way, we will see in the next section that in
the IR regularized system where some of the dimensions are
compact the classical result is indeed qualitatively
correct.

In the classical approximation the electric field as well as
the electric charge density in the vacuum vanish. Therefore
the dual vector potential has a ``pure gauge'' form
\begin{equation}
\chi_i=\partial_i\lambda
\end{equation}
As discussed earlier the dual potentials $\chi_i$ that are
given by the functions $\lambda$ with different values on
spatial infinity are not gauge equivalent.

Therefore in this approximation the QED vacuum is infinitely
degenerate with degeneracy that corresponds to global
transformations generated by $\Phi_S$. The expectation value
of a vortex operator in any of these vacua also does not
vanish and is given by
\begin{equation}
<V(C)>=\exp\{i\frac{2\pi}{e}\int dl_i\chi_i\}
\end{equation}

This still does not answer the question whether the
Goldstone theorem apply if $V(C)$ has a nonvanishing
expectation value. The problem is that $V(C)$ are nonlocal
operators and no local order parameters of $\Phi_i$ exist.
The situation with the electric charge in the Higgs phase
seems similar.  There the Goldsone theorem was not
applicable and no massless particle existed for the
following reason. Suppose one has a spontaneously broken
charge $Q$.  For the Goldstone theorem to hold
\cite{itzykson} there must exist an operator $O$ which
satisfies the following two conditions
\begin{equation}
\lim_{V\rightarrow\infty}<[Q_V,O]>\ne 0
\label{g1}
\end{equation}
\begin{equation}
\lim_{V\rightarrow\infty}<[\dot Q_V(t),O]>=0
\label{g2}
\end{equation}
Here $Q_V\equiv\int_Vd^DxJ_0(x)$ is the spontaneously broken
charge in the volume $V$.  To satisfy eq.(\ref{g1}) it is
sufficient to find any order parameter of $Q$ with
nonvanishing expectation value. It is less trivial to
satisfy the second equation.  If the operator $O$ is local,
then eq.(\ref{g2}) is satisfied automatically:
\begin{equation}
\lim_{V\rightarrow\infty}<[\dot
Q_V(t),O(x)]>=\lim_{V\rightarrow\infty}\int_{S:\partial V=S}dS^i
<[J^i(y,t),O(x)]>=0
\end{equation}
since in the limit $V\rightarrow\infty$ the points $x$ and
$y$ are infinitely far apart and the commutator vanishes for
any finite time $t$. However if $O$ is nonlocal, in general
it need not commute with $J_i$ at spatial infinity and
eq.(\ref{g2}) need not be satisfied.  This is indeed the
reason why the spontaneous breaking of electric charge is
not accompanied by an appearance of a Goldstone particle.

In the case of magnetic symmetry it turns out however, that
the Goldstone theorem is indeed applicable even though the
order parameter is nonlocal. To see this let us consider the
charge $\Phi_3$ - the magnetic flux in the $z$ direction.
The corresponding order parameter is $V(C_3)$ of
eq.(\ref{vi}). The regularized flux operator $\Phi_3(L)$ is
defined as in eq.(\ref{phii}) but without taking the limit
$L\rightarrow\infty$\footnote{One can also restrict the
integration in the perpendicular directions $ x_1$ and $x_2$
to a finite domain, but this is irrelevant to our argument}.
For any finite $L$, $V(C_3)$ is still an order parameter.
Therefore if $<V(C_3)>\ne 0$, eq.(\ref{g1}) is satisfied.
Furthermore, only the boundary of the volume $V$ in which
$\Phi_3(L)$ is defined which is perpendicular to the axis $
x_3$ is crossed by the fluxon created by $V(C_3)$. Therefore
the only nonvanishing contribution to eq.(\ref{g2}) can
arise from the commutator of $V(C_3)$ with the third
component of the magnetic current.  However, the third
component of the current is $\tilde F_{33}$ and vanishes
identically at all times due to antisymmetry of $\tilde
F_{\mu\nu}$. Therefore eq.({\ref{g2}) is satisfied also and
the Goldstone theorem is applicable. The corresponding
Goldstone boson is the linearly polarized photon with
magnetic field in the $x_3$ direction.

Clearly the same argument applies to all the charges
$\Phi_i$ if one chooses $V(C_i)$ as corresponding order
parameters.  In this way photons with any direction of the
wave vector and polarization should have a gapless
dispersion relation e.g. to be massless.

\subsection{Topological interpretation of electric charge.}

Let us now show that the electric charge has an explicit
topological interpretation in terms of the vortex operators.
First let us explain what do we mean by this.  As we
mentioned in the Introduction, the electric current is
trivially conserved.  In the usual representation of the
electrodynamics via gauge fields the charge however is not
explicitly given as some ``winding number'' but rather as a
surface integral of the electric field.

Consider a state with a pointlike charged particle at the
origin.  We know that if we place an infinite magnetic
vortex somewhere in space and move it adiabatically around
the charge, no matter how large the distance between the
vortex and the charge is, the Aharonov - Bohm ( or rather
the Aharonov - Casher) \cite{aharonov} phase will be
accumulated. In order for that to happen, the vortex must
complete the rotation around the charge. This means then,
that although locally the charged state at spatial infinity
is indistinguishable from the vacuum (for example
$F_{\mu\nu},J_\mu$ and $T_{\mu\nu}$ all vanish) there exists
some global characteristics which does distinguish between
them.

Remembering that the QED vacuum is in fact degenerate, the
natural possibility is that locally at every point at
infinity the charged state is similar to {\it one} of the
vacua, but moving from point to point we actually move from
one vacuum to another. If this is the case, then when we
complete the rotation we should, of course come back to the
same vacuum. If this closed path in the manifold of vacuum
states is incontractible there should be a topological
winding number associated with it. Given the fact that the
electric charge in QED has features characteristic of a
topological charge (trivially conserved and quantized) it is
natural to expect that it is identical to this winding
number.

Note that although the notion of topology of the manifold of
the vacua originates in the classical field theory, it has a
precise quantum meaning.  Suppose one has a vacuum
degeneracy in the quantum theory due to spontaneous breaking
of some symmetry group $G$. The different vacuum states will
differ not only in expectation value of the order parameter
$O$ but also all its correlators and, in fact all operators
which are nontrivial representations of $G$. However, since
the vacuum degeneracy is only due to the spontaneous
breaking of $G$, the VEV of any operator $O_i$ on which the
action of $G$ is free, unambigously determines the values of
all the other correlators.  Therefore the possible values of
this order parameter $O$ can be taken to parametrize the
manifold of the quantum vacua and the topology is identical
to the classical one.

Analogously one defines a notion of a topological soliton
which "interpolates between different vacuum states" at
spatial infinity. Usually (when the broken symmetry has a
local order parameter) this is the state with the following
properties.  Consider a chunk of space $T$ of linear
dimension $a$ at a distance $R$ from the soliton core, so
that $\frac{a}{R}\rightarrow 0$. Then the expectation value
of any local operator $O(x)$, $x\in T$ and any correlator of
local operators $O_1(x_1)...O_n(x_n)$, $x_1,...,x_n\in T$
will be equal to their vacuum expectation values in one of
the vacua. In another chunk $T_1$ which is also very far
from the soliton core but also far from $T$ so that
$|x_T-x_{T_1}|=o(R)$ the values of these correlators are
given by their expectation values in another vacuum state.
So that in this soliton state at each "point" at infinity
one has a vacuum state in the sence that all local and
quasilocal operators (operators with finite support) have
expectation values equal to their VEVs in a vacuum. These
vacua are however different at different points at infinity
and the mapping from the spatial boundary into the vacuum
manifold is not homotopic to a trivial map. The soliton
charge in this case is equal to the winding number
corresponding to the homotopy group $\Pi_{D-1}(M)$, where
$D-1$ is the dimension of a spatial boundary and $M$ is the
manifold of the vacua.

This was precisely the picture in QED in 2+1 dimensions. The
vacuum manifold was $S_1$ corresponding to the phase of the
VEV of the vortex operator $V(x)$.  In a charged state with
charge $n$ the configuration of the vortex field looked
asymptotically like a hedgehog :
$V(x)\rightarrow_{x\rightarrow \infty}e^{in\Theta}$, and the
electric charge was equal to the winding number
corresponding to the homotopy group $\Pi_1(S^1)=Z$.

The situation in $QED_{3+1}$ is slightly different. The
vacuum (at least in the classical approximation) is still
degenerate. However the broken symmetry group is represented
trivially on all local operators. The only operators that
carry the broken charges and whose VEVs therefore
distinguish between different vacua are the infinitely long
magnetic vortex lines $V(C)$. It is clear therefore, that in
any soliton like state (if it exists) all quasilocal
operators will have the same VEV at all points at spatial
infinity. The soliton is not characterized by $\Pi_2(M)$, or
rather $\Pi_2(M)=0$. To see the difference between different
regions of space far from the soliton core one has to
calculate the VEV of $V(C)$. One therefore has to divide the
spatial infinity not into quasipointlike regions but rather
into quasistringlike regions and compare VEVs of $V(C)$ and
their correlators (which fit into one such region).

The set of all the vortex operators is overcomplete. Like in
the case when a local order parameter exists, it is enough
to pick the minimal set of operators so that every group
element of the spontaneously broken group be represented
nontrivially. In our case the set of broken charges is
$\{\Phi_S\}$. The most convenient choice for $\{V(C)\}$ is
the set of all straight lines.

 The operators whose VEVs one compares should be
transformable into each other by {\it translation}. The
operation of translation does not change either the
orientation or the form of a string.  Moreover, all the
points on a string should be far from the soliton core.
Therefore for a given straight vortex line the set of
operators to which it should be compared can be chosen as
the set of all straight vortices having the same direction
and the same distance from the soliton core, in the limit
where this distance becomes infinite.  We see therefore that
the relevant homotopy is the first rather than the second
homotopy group $\Pi_1(M)$. As we have discussed earlier, the
vacuum manifold is infinitely dimensional, corresponding to
infinite number of the broken charges $\Phi_S$ and therefore
this homotopy group is huge.  However if we only consider
rotationally symmetric solitons, the things simplify
considerably. Since the straight lines in different
directions can be all transformed into each other by a
rotation, for the rotationally symmetric soliton the winding
numbers for all sets of straight lines is the same. Since
the vortex operator creating a straight line in a given
direction takes values in $S^1$, we see that for
rotationally symmetric configurations the soliton charges
must take values in $\Pi_1(S^1)=Z$.

Let us now calculate expectation values of the magnetic
vortex lines in the third direction in a state with electric
charge at the origin.  We again do this in the classical
approximation. The vortex operator which creates a fluxon
parallel to the third axis with coordinates $(X_1,X_2)$ is
\begin{equation}
V(X_1,X_2)=\exp\{i\frac{2\pi}{e}\int_{-\infty}^{\infty}
dx_3\int_{X_1}^\infty dx_1 E_2(x_1,X_2,x_3)\}
\end{equation}
In the classical approximation the phase factor is
proportional to the electric flux through the half plane
$(x_2=X_2, x_1>X_1)$.  For the spherically symmetric
configuration of a pointlike electric charge this is
proportional to the planar angle $\Theta$ between the vector
$(X_1,X_2)$ and the axis $x_1$. Since the total flux is
equal to $e$, we find
\begin{equation}
<V(R,\Theta)>=\exp\{i\Theta\}
\label{vsol}
\end{equation}
For pointlike electric charge this expression is valid for
any $R$.  If the charged state has some charge distribution,
the expression eq.(\ref{vsol}) will be still valid
asymptotically for $R\rightarrow\infty$. In the state with
electric charge $eN$, one clearly has
\begin{equation}
<V(R,\Theta)>=\exp\{iN\Theta\}
\label{vsolN}
\end{equation}

So we see that electrically charged states realize the
nontrivial windings of the vortex operators. The electric
charge is equal to the winding number.

 It is quite easy to construct states with winding numbers
corresponding to more general elements of $\Pi_1(M)$ and not
only $\Pi_1(S^1)$. For example consider a charged state
which is not sphericaly symmetic but has all the electric
flux lines asymptotically parallel to the ($x_1x_3$) plane.
In this case all the operators considered earlier will have
a unit expectation value, since no flux crosses the
($x_1x_3$) plane. However the fluxons in the direction
$x_2$, for example will still have a winding number $1$. So
this state has a nonzero winding with respect to
transformations generated by $\Phi_2$, but is trivial with
respect to transformations generated by $\Phi_3$.

However the mere fact that one can construct a state with a
given topological charge does not mean that it is
necessarily realized in the theory. It must also pass the
test of having a finite energy.  Electrically charged states
which are not asymptotically rotationally invariant have
infinite energy and are of no interest in QED.

As a final comment in this section we note that in the
classical approximation the Higgs phase can not be studied.
Since the vortex operator is defined as a unitary operator,
classically its VEV can not be zero and therefore we are
always in the Coulomb phase. This is similar to the
nonlinear $\sigma$ - model, where in the classical
approximation one does not see the unbroken phase.  Quantum
corrections of course induce the phase transition there. In
the present case as we will see in the following section,
the same phenomenon occurs.

\section{Quantum corrections and the infrared regularizarion.}
\subsection{Quantum corrections  to $<V(C)>$.}

We will now calculate the $<V(C_3)>$ taking into account the
lowest order quantum fluctuations.

Let us start with the Coulomb phase. The lowest order in $e$
correction to the classical result is
\footnote{Since the calculation is analogous to the corresponding
one in 2+1 dimensions we skip the details that can be found
in \cite{npb}}
\begin{equation}
<V(C_3)>=\exp\{-\frac{1}{2}(\frac{2\pi}{e})^2\int d^4k
a_i(k)a_j(-k)G_{ij}(k)\}
\label{vq}
\end{equation}
where
\begin{equation}
a_i(k)=\delta_{i2}\frac{1}{k_1}\delta(k_3)
\end{equation}
and $G_{ij}(k)$ is the propagator of the electric field
\begin{equation}
G_{ij}(k)
=i\{\frac{k_0^2}{k^2}
(\delta_{ij}-\frac{k_ik_j}{k_0^2})-\delta
_{ij}\}
\label{prope}
\end{equation}
The integral in eq.(\ref{vq}) is both ultraviolet and
infrared divergent.  Introducing the ultraviolet cutoff
$\Lambda$ and the infrared cutoff (in real space) L, we find
\begin{equation}
<V(C_3)>=\exp\{-(\frac{2\pi}{e})^2\Lambda L\}
\end{equation}

The reason for both divergencies is intuitively clear. The
ultraviolet divergence appears since the vortex line created
by $V(C)$ has zero thickness.  It can be dealt with by
either regularizing the vortex itself (making it finite in
cross section) or by multiplicative renormalization
\cite{polchinski}.
The infrared divergence comes about because of the infinite
length of the vortex line.

So we find that in one very important respect the quantum
corrections change the classical result. Now in the limit
$L\rightarrow\infty$ we have $<V(C)>\rightarrow 0$. The
situation is similar in some sense to 1+1 dimensional field
theories having a continuous global symmetry. There too, in
the classical approximation one can have nonvanishing order
parameter which, however is found to vanish when quantum
fluctuations are taken into account. As a result a
continuous symmetry is never broken in 1+1 dimensions.

There is however a very important physical difference
between the two cases.  In the 1+1 dimensional models the
order parameter is local. The vanishing of its expectation
value therefore persists also for a finite infrared cutoff
as long as the cutoff theory preserves the symmetry.
Technically, the system is disordered by the zero mode. If
the infrared regularization is done in such a way that the
"spontaneously broken" symmetry is not broken explicitely,
the zero mode is still present and it still leads to the
vanishing of the VEV of the order parameter. If the
regularization is such that the zero mode is given a finite
mass, the VEV can be nonzero, but the symmetry is then
broken explicitely.  In the case at hand though, the order
parameter is nonlocal and this nonlocality, rather than the
zero mode contribution is the factor which leads to the
vanishing of the VEV. This can be seen explicitely by taking
a massive rather than a massless propagator in
eq.(\ref{prope}). The result is still linearly infrared
divergent. Therefore it is clear that one can find an
infrared regularization which without explicit breaking of
the magnetic symmetry will yield a finite expectation value
for the vortex operator. In this regularized theory the
arguments conserning the Goldstone theorem and the
topological interpretation of the electric charge presented
in the previous section in the classical approximation will
be valid also in the full quantum theory.

But before doing that let us calculate $<V>$ in the Higgs
vacuum.  The most convenient way to do this is using the
euclidean path integral formalism.  The expectation value
$<V(C_3)>$ can be written in the following form
\cite{frohlich}:
\begin{equation}
<V(C_3)>=\int {\cal D} A_\mu {\cal
D}\phi\exp\{-[\frac{1}{4e^2}(\tilde F_{\mu\nu}-\tilde
f_{\mu\nu})^2+|D_\mu\phi|^2+U(\phi^*\phi)]\}
\end{equation}
where $$ \tilde f_{\mu\nu}(x)=\delta_{1[\mu }\delta_{\nu
]3}\delta(x_0)
\delta(x_2)\theta(x_1)$$ with $\theta(z)$ - a
step function.  For any given $x_3$ the field $\tilde f$
satisfies
\begin{equation}
\partial_\nu \tilde f_{\mu\nu}=\delta_{3\mu}\delta^3(x)
\end{equation}
If we now view $x_3$ as the euclidean time, $\tilde f$ is
the magnetic field of the Dirac string of a static magnetic
monopole propagating in time.  At the tree level therefore
the VEV is given by a Euclidean action of a static magnetic
monopole in the Higgs phase. In the Higgs phase magnetic
monopoles are linearly confined and the energy of a single
monopole diverges linearly with the dimension of a system.
The action therefore diverges quadratically and we obtain

\begin{equation}
<V(C_3)>=e^{-\alpha L_1L_3}
\end{equation}
where $\alpha$ is a dimensional constant, and $L_1$ and
$L_3$ are infrared cutoffs on the first and the third
directions respectively.

We see that the VEV vanishes much faster in the infrared
than in the Coulomb phase. In fact even if we make the
system finite in the direction of the vortex line, the VEV
still vanishes in the limit $L_1\rightarrow\infty$.  So even
though in the infinite system the VEV of the vortex operator
vanishes in both phases, there is a qualitative difference
in the dependence on the infrared cutoff. This difference
will be reflected in the VEV of a closed vortex loop
(t'Hooft loop). Evidently the large loops in the Coulomb
phase will have a perimeter law behaviour, while in the
Higgs phase - the area law.  This result of course coincides
with t'Hooft's discussion of expected behaviour of vortex
loops \cite{thooft1}.

\subsection{The infrared regularization.}

Let us now describe the simplest infrared regularized theory
which has a finite VEV of the vortex operator in the Coulomb
phase.  Consider QED defined on a spatial manifold which is
compact in the direction of the $x_3$ axis. This means that
all the gauge invariant fields ($B_i$, $E_i$, $J_0$ etc.)
must at all times obey the periodic boundary condition
\begin{equation}
O(x_1,x_2,x_3)=O(x_1,x_2,x_3+L)
\end{equation}

In this theory the magnetic flux $\Phi_3$ is still a
conserved charge. We will concentrate on it and on its order
parameter the vortex line $V(C_3)$.  As previously, $V(C_3)$
is a well defined operator, except that now the vortex line
it creates has a finite length $L$.  The calculation of
$<V(C_3)>$ is the same as previously. The only alteration is
that the photon's propagator must be modified according to
the new boundary condition, so that $k_3$ in
eq.(\ref{prope}) takes discrete values $k_3=
\frac{2\pi n}{L}$. The exact
form of the propagator, however does not
matter as we have seen earlier and we obtain
\begin{equation}
<V(C_3)>=e^{-\alpha\Lambda L}
\end{equation}

The proof of the Goldstone theorem goes through in precisely
the same way as in the unbounded case since the dual field
strength tensor remains antisymmetric. The Goldstone bosons
that appear due to spontaneous breaking of $\Phi_3$ are the
linearly polarized photons with magnetic field pointing in
the direction $ x_3$.

Note that in the Higgs phase $<V(C_3)>=0$ because of the
infinite extent of our system in the direction $ x_1$. The
Higgs - Coulomb phase transition therefore is attributed to
the spontaneous breaking of $\Phi_3$.

One also immediatelly realizes that an electrically charged
state has a nonzero winding of $V(C_3)$.  The configuration
of the electric field of a point charge near the location of
a charge is the same as in the unbounded case. However since
the electric flux can not escape through the boundary due to
periodic boundary conditions, near the boundary the electric
flux lines get squeezed and become parallel to the $x_1x_2$
plane, so that all the flux escapes to infinity (Fig.3.).
\begin{figure}
\caption{ The schematic distribution
 of the electric flux lines of the field of
a pointlike charge in a box of a finite
 hight $L$ with periodic boundary conditions.}
\label{f3}
\end{figure}

Repeating now the calculation of a previous section we find
that the surface associated with $V(C_3)$ collects the
electric flux proportional to the planar angle, and
therefore the eq.(\ref{vsol}) is still valid.

If we would have taken free instead of periodic boundary
conditions, part of the electric flux would have escaped
through the boundary and we wouldn't have had a complete
wind of $V(C_3)$. However in this case the electric charge
would also not be conserved, since charged particles would
be able to leave the system freely through the boundary. And
of course, if a charge is not conserved it can not be
topological.

Other infrared regularizations are possible. For example,
one could take the system to be finite in two directions $
x_3$ and $ x_2$. Then both $<V(C_3)>$ and $<V(C_2)>$ would
be nonvanishing in the Coulomb phase. The masslessness of
photons with two linear polarizations would then follow by
Goldstone's theorem. If the boundary conditions preserve the
$\frac{\pi}{2}$ rotations around the first axis, the finite
energy electrically charged states will carry a unit winding
of both $V(C_2)$ and $V(C_3)$.  Note however that the
regularization in which all three directions are made
compact is illegal since in this case $V(C)$ can not be
defined. The reason is that the surface of singularities
associated with $V(C)$ must be infinite and there are no
such surfaces in a completely finite system. The same is
true in $2+1$ dimensions where one can not define the local
vortex field $V(x)$ in a finite system with periodical
boundary conditions.

We see therefore, that any sensible infrared regularization
leads to a nonvanishing VEV of the vortex operators. The
Goldstone's theorem for the photon and the identity of the
electric charge with the winding number hold for any finite
value of the infrared cutoff. In this sense both results are
also true in the unbounded theory although the actual VEV of
the order parameter vanishes.

Note that $QED_4$ differs from a typical field theory in the
following respect. Usually the tendency of a smaller system
is towards a restoration of any broken symmetries, since the
potential barrier between the degenerate vacua becomes
smaller.  In a certain sense this is why the high
temperature phase is usually the one where all the
symmetries are restored.  In $QED_4$ as we have seen the
opposite happens. Due to the nonlocality of the order
parameter, its expectation values is actually larger for a
smaller system. This might be connected to the fact that in
$QED_4$ the high temperature phase is the Coulomb phase, in
which the flux symmetry is broken whereas the low
temperature, Higgs (superconducting) phase has the symmetry
restored.

\section{Discussion.}

In this paper we looked at QED from an unconventional point
of view. Instead of concentrating our attention on the
standard degrees of freedom like photons and charged
particles, we have analysed the behaviour of the dual
variables - the magnetic vortex lines. The picture that
transpires from this point of view is somewhat similar to
2+1 dimensional electrodynamics.

In the Coulomb phase the operators creating infinitely long
vortex lines $V(C)$ have ``finite expectation value per unit
length''.  What this means is that the expectation value of
such an operator in a system with a finite infrared cutoff
in the direction of the vortex line behaves as $e^{-\alpha
L}$ in the Coulomb phase.  The operators $V(C)$ are
eigenoperators of the magnetic symmetry generators $\Phi_S$.
Therefore in the Coulomb phase the magnetic symmetry group
is spontaneously broken.  Although the only order parameters
for $\Phi_S$ are the nonlocal vortex operators, the
Goldstone theorem is still applicable and the spontaneous
breaking of this symmetry leads to exact masslessness of the
photon.

The electric charge in this picture is topological and
corresponds to the homotopy group $\Pi_1(S^1)$ of possible
string configurations.

In the Higgs phase the VEV $<V(C)>$ vanishes. The magnetic
symmetry is restored and no massless excitations are
present. The Higgs - Coulomb phase transition is driven by
condensation of the magnetic vortices.

According to the standard lore a theory near a phase
transition (and also away from the phase transition but at
low energies) should be describable in terms of the Landau -
Ginzburg type Lagrangian for the order parameter. In the
case at hand this would not be a standard field theory but
rather a string theory of the vortex lines $V(C)$.  An
approximate derivation of this string theory is given in
\cite{kawai}.  In fact the dual field strength tensor
$\tilde F_{\mu\nu}$ can be expressed via $V(C)$ in the same
way as $F_{\mu\nu}$ is expressed via the Wilson line
\cite{migdal}
\begin{equation}
\tilde F_{\mu\nu}(x)
=V^\dagger\frac{\delta}{\delta S_{\mu\nu}(x)}
V(C)
\end{equation}
where $\frac{\delta}{\delta S_{\mu\nu}(x)}$ is the area
derivative at the point $x$.  The kinetic term therefore can
be rewritten in terms of the vortex creation operator as
\begin{equation}
\frac{\delta}{\delta S_{\mu\nu}}
V^\dagger (C)\frac{\delta}{\delta S_{\mu\nu}}
V(C)
\end{equation}

There is an interesting possibility that the exact dual
reformulation of QED$_4$ exists.  Then QED should be exactly
equivalent to an interacting string theory.  Clearly the
weakly interacting QED will be described by a strongly
interacting dual string theory. This must be so, since the
spectrum of a free string theory contains an infinite number
of particles, whereas the spectrum of QED contains just the
familiar excitations: the photon and the charged particles.
It should also be noted that a massless photon will arise in
this string theory in a way very different from massless
gauge particles in a free string theory, since it is
massless in the phase in which the strings are condensed.

There are many further questions which have been asked in
the context of 2+1 dimensional gauge theories which we did
not address in this paper.  For example, what elements of
the dual picture should be modified if the matter fields are
fermionic. But the most interesting one is, perhaps can this
picture be generalized to nonabelian theories. It would be
very rewarding to have a simple qualitative picture of
confinement based on topological interpretation of electric
charge similar to the one available in 2+1 dimensions
\cite{ym}.  There constituent quarks can be understood as
topological defects like electrric charges in QED, but the
flux symmetry is broken explicitly (by the nonperturbative
monopole instanton effects). As a result of this explicit
breaking the vacuum is nondegenerate (or has a finite
degeneracy) and the topological defects are linearly
confined.  In nonabelian theories in 3+1 dimensions there
are also nonperturbative effects due to magnetic monopoles
(which now are particles rather than instantons). The
appearance of the monopoles again breaks explicitely the
magnetic symmetry since the dual field strength is not
conserved anymore.  It is interesting to see whether this
explicit breaking leads to linear confinement of the
topological defects as in 2+1 dimensions, although the
defects now are of quite a different nature. It is also
worth noting that in $SU(N)$ theories with adjoint matter
fields only, the monopoles carry $N$ units of the elementary
Dirac quantum. Therefore the discreet subgroup of the
magnetic group will still survive (just like in 2+1
dimensions). The phase transitions between the "completely
broken" Higgs phase and a confinement phase can then be
attributed to the spontaneous breaking of this discreet
symmetry.

\end{document}